# Development of Low Temperature and High Magnetic Field X-Ray Diffraction Facility

Aga Shahee, Shivani Sharma, K. Singh, N. P. Lalla[*] and P. Chaddah

*UGC-DAE Consortium for Scientific Research, University campus, Khandwa road Indore-452001*

**Abstract**

The current progress of materials science regarding multifunctional materials (MFM) has put forward the challenges to understand the microscopic origin of their properties. Most of such MFMs have magneto-elastic correlations. To investigate the underlying mechanism, it is therefore essential to investigate the structural properties in the presence of magnetic field. Keeping this in view low temperature and high magnetic field (LTHM) powder x-ray diffraction (XRD), a unique state-of-art facility in India has been developed at CSR Indore. This setup works on symmetric Bragg Brentano geometry using a parallel incident x-ray beam from a rotating anode source working at 17 kW. Using this one can do structural studies at non-ambient conditions i.e. at low- temperatures (2-300 K) and high magnetic field (+8 to -8 T). The available scattering angle ranges from 5° to 115° 2θ with a resolution better than 0.1°. The proper functioning of the setup has been checked using Si sample. The effect of magnetic field on the structural properties has been demonstrated on $Pr_{0.5}Sr_{0.5}MnO_3$ sample. Clear effect of field induced phase transition has been observed. Moreover, the effect of zero field cooled and field cooled conditions is also observed.

Corresponding author:nplallaiuc82@gmail.com

# INTRODUCTION

The study on multifunctional materials (MFM) is an interdisciplinary field and has strong impact on the technological as well as basic understanding of the materials. The phenomenon of phase transition is an important issue in materials. Till date temperature (T) and pressure (P) has been the only thermodynamic control variables to investigate it. But for magnetic materials [1, 2], the applied magnetic field also has been shown as a thermodynamic variable. Keeping in view the emergence of MFMs, such observations [1, 2] indicate the importance of in-field structural studies of the materials i.e. performing x-ray diffraction (XRD) in the presence of magnetic field. Such facilities are available at synchrotron/neutron sources. But very few such setup on Lab. sources are available worldwide [3, 4], none in India. Such state-of-the art facility is essential for condensed matter research. Such a facility has been developed at UGC-DAE CSR Indore, India.

# INSTRUMENTATION

Keeping the above motivation in view such a low temperature (LT) high magnetic field (HM) powder XRD setup has been developed around an 18 kW rotating anode source (Rigaku) at UGC-DAE CSR Indore,

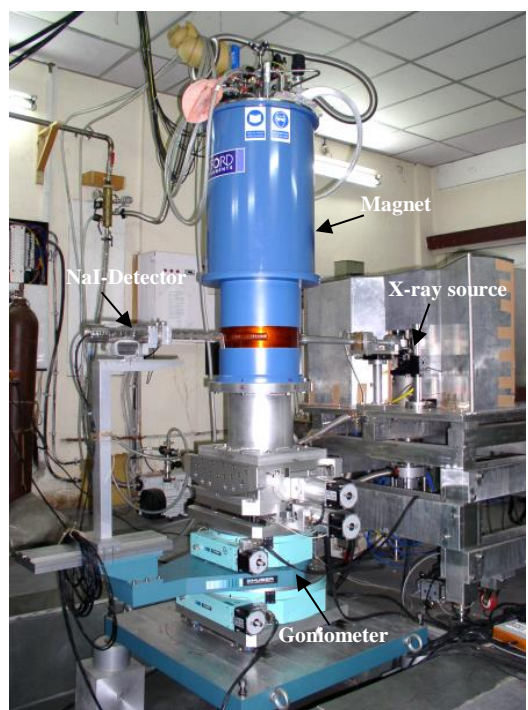

**FIGURE 1.** A photographic view showing integration of various components used in the LTHM XRD set up (unshielded).

India. The LTHM XRD uses 8T split-pair superconducting magnet (Oxford) which is mounted on a heavy goniometer (Huber) equipped with all necessary motions along with data collection accessories. For minimizing the direct effect of stray field on the source the magnet is kept at a distance of ~1meter. The diverging vertical line source coming out of the x-ray tube is focused and made parallel using a parabolic mirror (Xenox). The setup needs only ~0.3cc of sample which is mounted between magnets through a specially designed insert. The scattered x-ray is detected using NaI detector. The whole setup is tightly shielded for the scattered x-rays using a lead hutch. A photographic view of the setup is shown in figure 1. To control all motions for alignment and data collection a computer programme was developed. The available scattering angle ranges from 5° to 115° of 2θ with resolution better than 0.1°. We can collect statistically very good XRD data within an hour from 10 to 110° 2θ with peak to background ratio ~ 500. To check the proper functioning of the goniometer and the possible artifacts caused due to effect of stray-field on goniometer, source and detector; room temperature (RT) and LTHM XRD of Si powder was carried out. To demonstrate field induced structural changes LTHM XRD of $Pr_{0.5}Sr_{0.5}MnO_3$ (PSMO) was carried out in zero field cool (ZFC) and field cool (FC) conditions.

## RESULTS AND DISCUSSION

A Reitveld refined pattern with $\chi^2 = 2.53$ of Si at RT as shown in figure 2(a) very well matches with the reported data and therefore approves the proper functioning of the goniometer and the computer programme developed for alignment/data collection. The absence of any peak shift in the XRD patterns of Si taken in the presence of magnetic field, as shown in figure 2(b), indicates the absence of any artifact due to stray-field. For further clarifications the insets of figure 2b show the zoomed view of low and high 2θ peaks. The figure 2(c) shows the field induced structural changes. Transition of LT antiferromagnetic (AFM) *Fmmm* phase into high temperature ferromagnetic (FM) *I*4/*mcm* in PSMO takes place as a function of applied field at 4.2 K. Further, figure 2(d) shows the comparative increase of FM *I*4/*mcm* phase fraction in FC conditions as compared to ZFC.

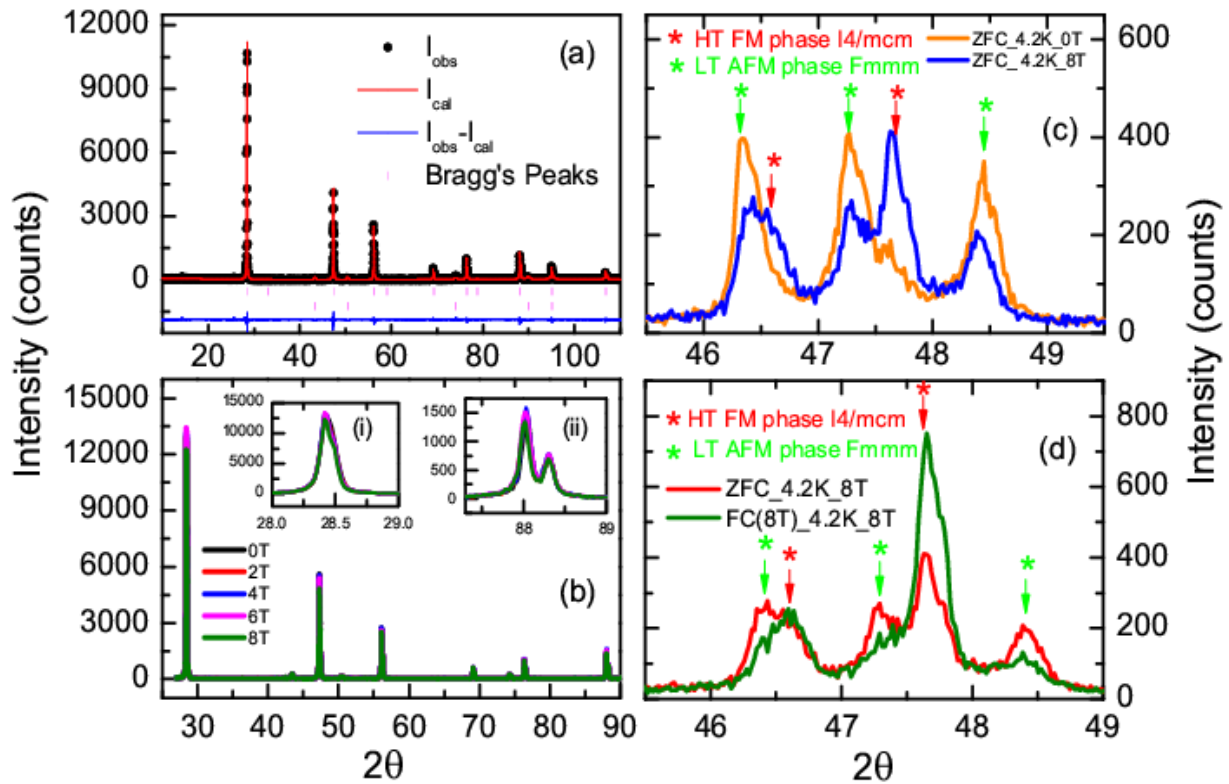

**FIGURE 2.** (a) Rietveld refined room temperature XRD pattern of Si powder taken at 10 kW. (b) XRD patterns of Si powder taken at different magnetic fields at 96 K at 12 kW. (c) Low temperature high magnetic fields XRD patterns of $Pr_{0.50}Sr_{0.50}MnO_3$ under zero field cooling condition collected in zero field and at 8T magnetic field, respectively. (d) Data taken at 4.2 K and 8T but with different histories, viz. ZFC and FC, of reaching that H,T point.

## CONCLUSION

A Lab. source based LTHM XRD setup has been developed. The applicability of this setup for field induced structural studies has been successfully demonstrated by showing the occurrence of field induced phase transition of LT AFM *Fmmm* phase into high temperature FM *I*4/*mcm* phase in PSMO.

## ACKNOWLEDGEMENT

Authors would like to thank Director Dr. A. K. Sinha, Centre Director Dr. V. Ganesan and Former Centre Director Prof. A. Gupta for their constant support and encouragements to develop this facility. Authors also like to thank Dr. A. Banerjee for useful discussion regarding magnetic shielding. NPL and KS are very thankful to Dr. R. Rawat, Er. P. Sarvanan and Ms. Preeti Bhardwaj for their frequent help during magnet cooling and also thankful to Mr. N. L. Ghodke for technical help regarding hardware/software for x-ray shutter control. AS would like to thank CSIR, New Delhi, for providing financial support in the form of SRF.